\documentclass[english,conference]{IEEEtran}
\usepackage[T1]{fontenc}
\usepackage[latin9]{inputenc}
\usepackage{amsthm}
\usepackage{amsmath}
\usepackage{amssymb}
\usepackage{graphicx}
\PassOptionsToPackage{normalem}{ulem}
\usepackage{ulem}

\makeatletter
\theoremstyle{plain}
\newtheorem{thm}{\protect\theoremname}
\theoremstyle{definition}
\newtheorem{example}[thm]{\protect\examplename}

\ifCLASSINFOpdf
\else
\fi
\makeatother

\usepackage{babel}
\providecommand{\examplename}{Example}
\providecommand{\theoremname}{Theorem}

\begin{document}

\title{A hybrid TIM-NOMA scheme for the SISO Broadcast Channel}

\author{\IEEEauthorblockN {Vaia Kalokidou} \IEEEauthorblockA{Communication
Systems and\\Networks Research Group\\MVB, School of Engineering\\University
of Bristol, UK\\ Email: eexvk@bristol.ac.uk}\and\IEEEauthorblockN
{Oliver Johnson} \IEEEauthorblockA{ Department of Mathematics\\University
of Bristol, UK\\ Email: o.johnson@bristol.ac.uk}\and \IEEEauthorblockN{Robert
Piechocki} \IEEEauthorblockA{Communication Systems and\\ Networks
Research Group\\MVB, School of Engineering\\University of Bristol,
UK\\ Email: r.j.piechocki@bristol.ac.uk}}
\maketitle
\begin{abstract}
Future mobile communication networks will require enhanced network
efficiency and reduced system overhead due to their user density and
high data rate demanding applications of the mobile devices. Research
on Blind Interference Alignment (BIA) and Topological Interference
Management (TIM) has shown that optimal Degrees of Freedom (DoF) can
be achieved, in the absence of Channel State Information (CSI) at
the transmitters, reducing the network's overhead. Moreover, the recently
emerged Non-Orthogonal Multiple Access (NOMA) scheme suggests a different
multiple access approach, compared to the current orthogonal methods
employed in 4G networks, resulting in high capacity gains. Our contribution
is a hybrid TIM-NOMA scheme in Single-Input-Single-Output (SISO) $K$-user
cells, in which users are divided into $T$ groups, and $1/T$ DoF
is achieved for each user. By superimposing users in the power domain,
we introduce a two-stage decoding process, managing ``inter-group''
interference based on the TIM principles, and ``intra-group'' interference
based on Successful Interference Cancellation (SIC), as proposed by
NOMA. We show that for high SNR values the hybrid scheme can improve
the sum rate by at least 100\% when compared to Time Division Multiple
Access (TDMA).
\end{abstract}

\section{Introduction}


Future increase in the number of mobile devices, using data-hungry
applications, will lead to highly dense cellular networks, demanding
high capacity performance with the least possible system overhead.
Interference Alignment (IA), introduced by Maddah-Ali, Motahari and
Khandani in {[}1{]} and Cadambe and Jafar in {[}2{]}, allows in the
$K$-user interference channel $K/2$ Degrees of Freedom (DoF) to
be achieved, assuming global perfect CSI. IA differs from other interference
management schemes, as it attempts to align interference, rather than
avoid, reduce or cancel it. 

However, IA requirement of full CSI is infeasible and costly. The
scheme of Blind IA (BIA), presented by Wang, Gou and Jafar in {[}3{]}
and Jafar in {[}4{]}, for certain network scenarios, can achieve full
DoF in the absence of CSI at the transmitters (CSIT), reducing considerably
the system overhead. Additionally, in {[}5{]} Jafar introduces how
the BIA scheme can be employed in certain cellular networks, including
heterogeneous networks, by seeing frequency reuse as a simple form
of IA. In {[}6{]}, Jafar introduces the Topological Interference Management
(TIM) scheme, which can be considered as a form of BIA in which the
position of every user in the cell(s), and therefore the strength
of their channels, is taken into account. Requiring only knowledge
of the network's topology at the transmitters, 1/2 DoF can be achieved
for every user in the SISO Broadcast Channel (BC), by treating weak
interference links as noise. Moreover, in {[}7{]} Sun and Jafar discuss
the implications of increasing the number of receive antennas resulting
in an increase on the network's DoF.

In {[}8{]}, Saito et al. propose a Non-Orthogonal Multiple Access
(NOMA) scheme for future radio access, in contrast to the Ortogonal
Frequency Division Multiple Access (OFDMA) and Single Carrier-Frequency
Division Multiple Access (SC-FDMA) orthogonal schemes currently adopted
by 4G mobile systems. According to the NOMA scheme, multiple users
are superimposed in the power domain and Successful Interference Cancellation
(SIC) reception is performed at the decoding stage, ultimately improving
the capacity and throughput performance. Furthermore, Benjebbour et
al. in {[}9{]} present the benefits of NOMA and discuss its performance
considering adaptive modulation and coding, and frequency-domain scheduling.
Moreover, Ding and co-authors in {[}10{]}-{[}11{]} discuss the superior
performance of NOMA in terms of ergodic sum rates and the importance
of power allocation, and a cooperative NOMA scheme where users with
higher channel gains have prior information about other users' messages,
respectively. Finally, Ding, Fan and Poor in {[}12{]} study user pairing
on two NOMA schemes and how it affects the sum rate. The first scheme,
F-NOMA, with fixed power allocation, pairs users with very distinctive
channel conditions, whereas the second one, CR-NOMA, inspired by cognitive
radio, pairs users with similar channel conditions.

In this paper, based on {[}6{]} and {[}8{]}, we introduce a hybrid
TIM-NOMA scheme in general $K$-user SISO cells. Our contribution
is the combination of the TIM and NOMA schemes, in a two-stage decoding
way, dividing users in $T$ groups. In the first-stage, we apply the
TIM scheme to manage ``inter-group'' interference, with no need
to ignore weak interference links. In the second-stage, we employ
NOMA, at every group of users separately, to manage ``intra-group''
interference through SIC. Finally, we discuss how the sum rate performance
of the system is significantly improved with the employment of the
hybrid scheme when compared to Time Division Multiple Access (TDMA).

The rest of the paper is organized as follows. Section II presents
the general description of the hybrid scheme, with the aid of an example
model with $K=5$ users, including the determination of the transmit
power, and the two-stage decoding process. Section III presents the
achievable rate formula for every user in the network. Finally, Section
IV gives an overview of our results, illustrated with graphs, discussing
how the users' distance from the basestation, and the amount of interference
they end up considering as noise affects their performance.

\section{System Model}

Consider the Broadcast Channel (BC) network, as shown in Figure 1,
for $K=5$ users. At the SISO BC of the cell, there is one transmitter
$T_{x}$ with 1 antenna, and $K$ users equipped with 1 antenna each.
Transmitter $T_{x}$ has 1 message to send to every user, and moreover,
when it transmits to user $k$, where $k\in\left\{ 1,2,...,K\right\} $,
it causes interference to all the other $K-1$ users in the macrocell.
The radius of the cell is considered as $R=5$ $km$, and the distance
of every user from the basestation is given by $d_{k}$. 

Furthemore, users are divided into $T$ groups $\left\{ G_{1},G_{2},...,G_{T}\right\} $
in such a way so that there are always $T-1$ users from the remaining
$T-1$ groups separating 2 users from the same group, and to place
users with considerable difference in their channel strengths in the
same group. The operation is performed over $T$ time slots, over
which we assume that channel coefficients remain the same. The transmitter
has only knowledge of the topology of the network.

According to the NOMA scheme, described fully in {[}8{]}-{[}9{]},
users are multiplexed, in the power domain, at the transmitters, and
then at the receivers, signal separation is performed based on SIC.
Decoding is performed based on an optimal order (in the order of descreasing
channel gains divided by the power of noise and interference), resulting
in every user being able to decode the signals of users coming before
them in the decoding order.

The general concept of the hybrid TIM-NOMA scheme, is that every user,
in order to recover its desired signal, uses the principles of TIM
to manage interference coming from transmissions to users NOT belonging
to their own group (i.e. their channel strengths are quite similar),
and the principles of NOMA to manage interference due to transmissions
to users belonging in their own group (i.e. their channel strengths
are quite different). 

According to our research, NOMA seems to work better when applied
to users with considerable difference in their channel gains. Therefore,
introducing TIM in the NOMA scheme, and splitting users into groups,
provides a solution for the cases where users' gains do not differ
much. The aforementioned reason, combined with fact that both schemes
do not require CSIT, as discussed in {[}6{]} and {[}9{]}, results
in a very smooth and successful combination of them.

In this paper, we will use an example model, to present the hybrid
TIM-NOMA scheme, where we consider $K=5$ users, $T=2$ time slots
and groups $\left\{ G_{1},G_{2}\right\} $. Users $1$,$3$ and $5$
are in group $G_{1}$, and users $2$ and $4$ are in group $G_{2}$.
Finally, the users' distances from the transmitter are given by: $d_{1}=0.5$
$km$, $d_{2}=1.5$ $km$, $d_{3}=2.5$ $km$, $d_{4}=3.5$ $km$,
$d_{5}=4.5$ $km$.

\begin{figure}
\begin{centering}
\includegraphics[width=0.9\columnwidth]{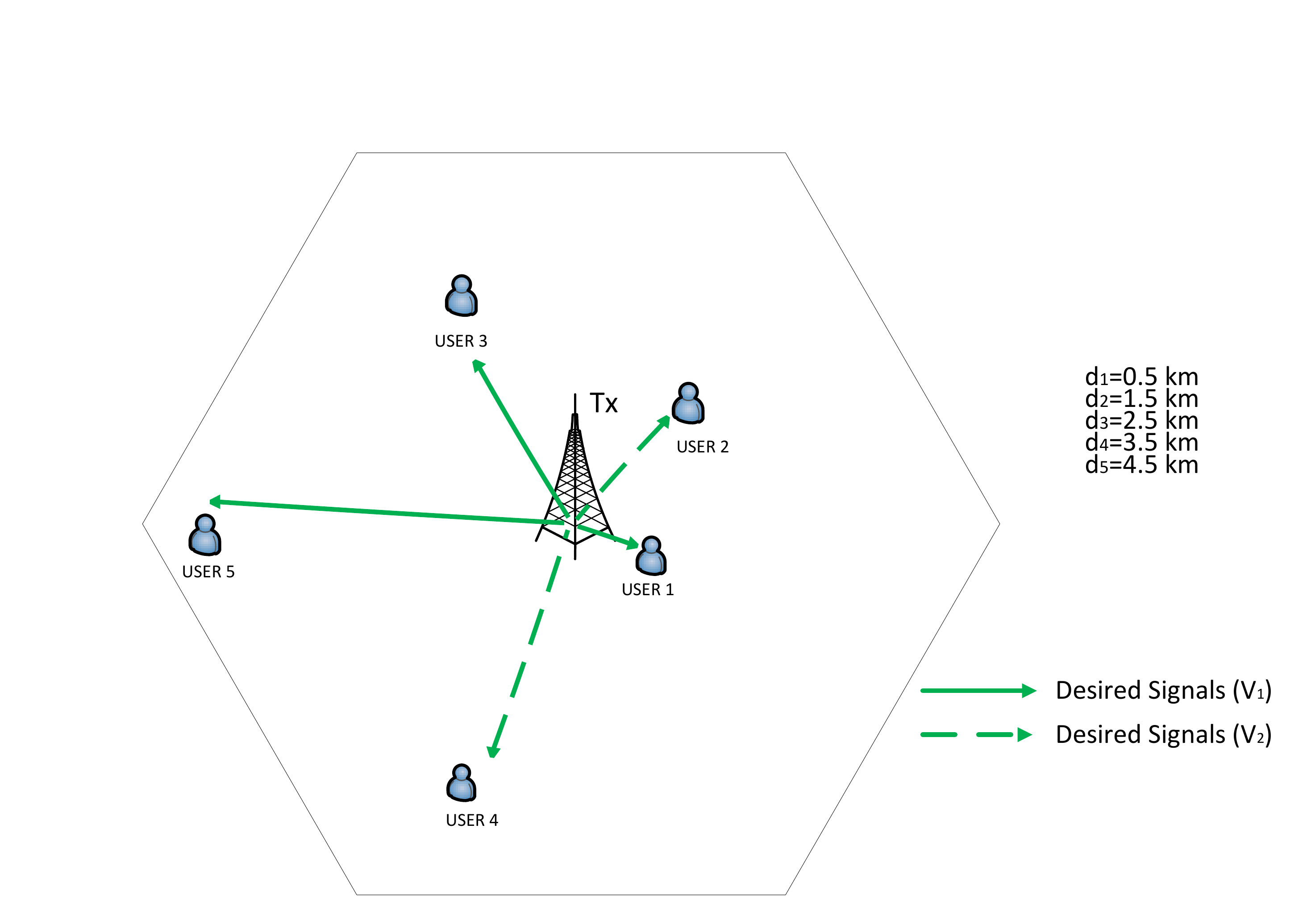}
\par\end{centering}

\caption{The hybrid NOMA-TIM scheme for a $K=5$ users cell. Users are divided
into $T=2$ groups: group $G_{1}$ (aligned along $\mathbf{V}_{1}$)
with users $1,3,5$ and group $G_{2}$ (aligned along $\mathbf{V}_{2}$)
with users $2,4$.}
\end{figure}

\subsection{Transmitted Power}

The $T\times1$ signal at receiver $k$ is given by:

\begin{equation}
\mathbf{y}_{k}=\mathbf{H}_{k}\mathbf{x}+\mathbf{z}_{k}
\end{equation}

Due to the users' different locations, channel coefficients are statistically
independent, and follow an i.i.d. Gaussian distribution $\mathcal{CN}(0,1)$.
$\mathbf{H}_{k}\in\mathbb{C}^{T\times T}$ is the channel transfer
matrix from $T_{x}$ to receiver $k$ and is given by $\mathbf{H}_{k}=\sqrt{\gamma_{k}}\left(\mathbf{I}_{T}\otimes h_{k}\right)$,
(here and throughout $\otimes$ denotes the Kronecker (Tensor) product),
with $h_{k}$ denoting the channel coefficient from $T_{x}$ to $k$
for one time slot. Moreover, $\gamma_{k}=\frac{1}{d_{k}^{n}}$ denotes
the path loss, and $n$ is the path loss exponent considered for an
urban environment, i.e. $n=3$. Finally, $\mathbf{z}_{k}\sim\mathit{\mathcal{CN}}(0,\sigma_{n}^{2}\mathbf{I}_{T})$
denotes independent Additive White Gaussian Noise (AWGN) at the input
of receiver $k$.

Taking into consideration the position of each user $k$ in the cell,
and therefore its distance $d_{k}$ from the basestation, ordering
users increasingly, in terms of $d_{k}$, the following relationship,
regarding their channel gains normalized by the noise power (assuming
the same noise power for all receivers), follows:

\begin{equation}
\frac{\left|h_{1}\right|^{2}}{\sigma_{n}^{2}}>\frac{\left|h_{2}\right|^{2}}{\sigma_{n}^{2}}>...>\frac{\left|h_{K-1}\right|^{2}}{\sigma_{n}^{2}}>\frac{\left|h_{K}\right|^{2}}{\sigma_{n}^{2}},
\end{equation}

with user $1$ being very close to the basestation and user $K$ at
the edge of the cell. Therefore, weaker channels, of users' being
far from the basestation, need to be boosted, such that the following
holds for their transmit power:

\begin{equation}
P_{K}>P_{K-1}>...>P_{2}>P_{1}
\end{equation}

The energy of the input symbol $x_{k}\in\mathbb{C}$, of each user
$k$, is defined as:

\begin{equation}
\mathrm{E}\left[\left|x_{k}\right|^{2}\right]=1
\end{equation}

For every user $k$ in the cell, we choose to take its transmitted
power given by:
\begin{equation}
P_{k}=a^{2}\frac{d_{k}^{2}}{\sum_{j=1}^{K}d_{j}^{2}},
\end{equation}

where $a\in\mathbb{R}$ is a constant determined by power considerations.
The total transmit power is given by the power constraint:
\begin{equation}
P_{T}=\left(\sum_{j=1}^{K}P_{j}\right)\mathrm{norm}(x_{k})=a^{2}
\end{equation}

\subsection{Stage 1 - ``Inter-group'' interference management -Topological
Interference Management (TIM) scheme}

In the network, there will be $T$ precoding vectors $\mathbf{v}_{t}$,
where $t\in\left\{ 1,2,...,T\right\} $, which are $T\times1$ unit
vectors. The choice of precoding vectors, carrying messages to users
in the cell, is not unique, and we choose them in such a way so that
every precoding vector $\mathbf{v}_{t}$ is orthogonal to all the
remaining $T-1$ precoding vectors.

The $T\times1$ transmitted vector $\mathbf{x}$ is given by:
\begin{equation}
\mathbf{x}=\sum_{k=1}^{K}\sqrt{P_{k}}\mathbf{v}_{t(k)}x_{k},
\end{equation}

with $t(k)\in\left\{ 1,2,...,T\right\} $ denoting the number of the
group $G_{t}$ each user $k$ belongs to.
\begin{example}
For the example model, the precoding vectors $\mathbf{v}_{1}$ and
$\mathbf{v}_{2}$, for groups $G_{1}$ and $G_{2}$ respectively,
are given by:
\end{example}
\begin{equation}
\mathbf{v}_{1}=\begin{bmatrix}1/2\\
\sqrt{3}/2
\end{bmatrix}
\end{equation}

\begin{equation}
\mathbf{v}_{2}=\begin{bmatrix}-\sqrt{3}/2\\
1/2
\end{bmatrix},
\end{equation}

and the $2\times1$ transmitted vector is:
\begin{equation}
\mathbf{x}=\sum_{k=1}^{5}\sqrt{P_{k}}\mathbf{v}_{t(k)}x_{k},
\end{equation}

where for $G_{1}=\left\{ 1,3,5\right\} $ and $G_{2}=\left\{ 2,4\right\} $.

As a result of the way precoding vectos are determined, receivers
of the same group $G_{i}$ see their desired signals along $\mathbf{v}_{i}$,
and undesired signals from users not in their group along remaining
$T-1$ precoding vectors $\left\{ \mathbf{v}_{j}\right\} $, for $j=1,...,T$
and $j\neq i$. Managing interference coming from transmissions from
users not belonging in their own group, and based on the example given
in {[}6, Section 4{]}, every receiver $k$, of the same group $G_{i}$,
can partially recover their signal by projecting their received signal
$\mathbf{y}_{k}$ along $\mathbf{v}_{i}$, which by definition is
orthogonal to all the other $T-1$ precoding vectors $\left\{ \mathbf{v}_{j}\right\} $,
for $j=1,...,T$ and $j\neq i$. 

\begin{flushleft}
\textbf{Theorem }1: \emph{Multiplying the received signal $\mathbf{y}_{k}$
with the transpose of the precoding vector $\mathbf{v}_{i}$, the
resulting signal at every receiver $k$, is given by:}
\par\end{flushleft}

\[
\begin{gathered}\begin{split}\widetilde{\mathbf{y}_{k}} & =\mathbf{v}_{i}^{T}\mathbf{H}_{k}\left(\sum_{j\in G_{i}}\sqrt{P_{j}}\mathbf{v}_{i}x_{j}\right)+\widetilde{\mathbf{z}_{k}}\end{split}
\end{gathered}
\]

\emph{
\begin{equation}
\begin{split} & =\sqrt{\gamma_{k}}h_{k}\left(\sum_{j\in G_{i}}\sqrt{P_{j}}x_{j}\right)+\widetilde{\mathbf{z}_{k}},\end{split}
\end{equation}
}

\emph{where $k\in G_{i}$, and $\widetilde{\mathbf{z}_{k}}=\mathbf{v}_{i}^{T}\mathbf{z}_{k}$
remains white noise with the same variance.}
\begin{IEEEproof}
We show that $\mathbf{v}_{i}^{T}$ removes ``inter-group'' interference,
i.e. interference resulting from transmissions to users in groups
$\left\{ G_{j}\right\} $ for $j=1,...,T$ and $j\neq i$, at the
\emph{k}th receiver:
\[
\mathbf{v}_{i}^{T}\mathbf{y}_{k}=\mathbf{v}_{i}^{T}\left(\sqrt{\gamma_{k}}\left(\mathbf{I}_{T}\otimes h_{k}\right)\sum_{k=1}^{K}\sqrt{P_{k}}\mathbf{v}_{t(k)}x_{k}+\mathbf{z}_{k}\right)
\]

\begin{equation}
=\mathbf{v}_{i}^{T}\sqrt{\gamma_{k}}\left(\mathbf{I}_{T}\otimes h_{k}\right)\left(\sum_{j\in G_{i}}\sqrt{P_{j}}\mathbf{v}_{i}x_{j}\right)+\mathbf{W}_{i}^{T}\mathbf{z}_{k}
\end{equation}
where by definition, for $j=1,...,T$ and $j\neq i$, $\mathbf{v}_{i}^{T}\mathbf{v}_{j}=0$.\end{IEEEproof}
\begin{example}
For the example model, for groups $G_{1}$ and $G_{2}$ respectively:
\end{example}
\begin{equation}
\mathbf{v}_{1}^{T}=\begin{bmatrix}\frac{1}{2} & \frac{\sqrt{3}}{2}\end{bmatrix}
\end{equation}

\begin{equation}
\mathbf{v}_{2}^{T}=\begin{bmatrix}\frac{-\sqrt{3}}{2} & \frac{1}{2}\end{bmatrix}
\end{equation}

The $1\times1$ post-processed signals at receivers are:

for $i=1,3,5$:

\begin{equation}
\widetilde{\mathbf{y}_{i}}=\mathbf{v}_{1}^{T}\mathbf{H}_{i}\left(\sum_{j=1,3,5}\sqrt{P_{j}}\mathbf{v}_{1}x_{j}\right)+\mathbf{v}_{1}^{T}\mathbf{z}_{1}
\end{equation}

and for $i=2,4$:

\begin{equation}
\widetilde{\mathbf{y}_{i}}=\mathbf{v}_{2}^{T}\mathbf{H}_{i}\left(\sum_{j=2,4}\sqrt{P_{j}}\mathbf{v}_{2}x_{j}\right)+\mathbf{v}_{2}^{T}\mathbf{z}_{2}
\end{equation}

\subsection{Stage 2 - ``Intra-group'' interference management - Non-Orthogonal
Multiple Access (NOMA) scheme}

The concept of NOMA will be applied in each group $G_{t}$ separately.
Based on {[}8, Section 3{]}, for every group $G_{t}$, the SIC process
is applied at every receiver. All users are ordered increasingly by
their channel gain $\left|h_{k}\right|^{2}$ normalized by the noise
power $\sigma_{n}^{2}$. Each user $k$ can correctly decode the signals
of users, in their own group, whose channel gain by noise power ratio
is smaller than theirs, i.e. come before them in (2), by considering
their own signal as noise. In the case where user $k$ receives interference
from transmissions to users in their own group that have a higher
channel gain by noise power ratio than they do, then user $k$ simply
decodes its own signal considering ``intra-group'' interference
from users, in their own group, who come after them in (2), as noise.
Maximum-Likelihood (ML) reception is performed every time a user decodes
its own or another user's signal.
\begin{example}
For the example-model, the decoding order for the users is:
\end{example}
\begin{equation}
\frac{\left|h_{1}\right|^{2}}{\sigma_{n}^{2}}>\frac{\left|h_{2}\right|^{2}}{\sigma_{n}^{2}}>\frac{\left|h_{3}\right|^{2}}{\sigma_{n}^{2}}>\frac{\left|h_{4}\right|^{2}}{\sigma_{n}^{2}}>\frac{\left|h_{5}\right|^{2}}{\sigma_{n}^{2}}
\end{equation}

\uline{In group $G_{1}$:} Receiver $5$ decodes its own signal,
considering interference from transmissions to users $1$ and $3$
as noise.

Receiver $3$ decodes first signal $x_{5}$ (finding $\widetilde{x_{5}}$),
considering its own signal as noise, and substracts the estimate $\widetilde{x_{5}}$
from its post-processed signal $\widetilde{\mathbf{y}_{3}}$. Then,
it decodes its own signal, considering interference from transmissions
to user $1$ as noise:

\begin{equation}
\widetilde{\widetilde{\mathbf{y}_{3}}}=\left(\widetilde{\mathbf{y}_{3}}-\mathbf{v}_{1}^{T}\sqrt{\gamma_{3}}\left(\mathbf{I}_{T}\otimes h_{3}\right)\mathbf{v}_{1}\widetilde{x_{5}}\right)
\end{equation}

Receiver $1$ decodes first signal $x_{5}$ (finding $\widetilde{x_{5}}$)
and then $x_{3}$ (finding $\widetilde{x_{3}}$) , substracting every
time the estimate of the interfering signal from its post-processed
one, considering its own signal as noise, eventually decoding its
own, interference-free, signal:

\begin{equation}
\widetilde{\widetilde{\mathbf{y}_{1}}}=\left(\widetilde{\mathbf{y}_{1}}-\mathbf{v}_{1}^{T}\sqrt{\gamma_{1}}\left(\mathbf{I}_{T}\otimes h_{1}\right)\mathbf{v}_{1}\widetilde{x_{5}}\right)-\mathbf{v}_{1}^{T}\sqrt{\gamma_{1}}\left(\mathbf{I}_{T}\otimes h_{1}\right)\mathbf{v}_{1}\widetilde{x_{3}}
\end{equation}

\uline{In group $G_{2}$:} Receiver $4$ decodes its own signal,
considering interference from transmissions to user $2$ as noise.

Finally, receiver $2$ decodes first signal $x_{4}$ (finding $\widetilde{x_{4}}$),
considering its own signal as noise, and substracts the estimate $\widetilde{x_{4}}$
from its post-processed signal $\widetilde{\mathbf{y}_{2}}$. Then,
it decodes its own signal:

\begin{equation}
\widetilde{\widetilde{\mathbf{y}_{2}}}=\left(\widetilde{\mathbf{y}_{2}}-\mathbf{v}_{2}^{T}\sqrt{\gamma_{2}}\left(\mathbf{I}_{T}\otimes h_{2}\right)\mathbf{v}_{2}\widetilde{x_{4}}\right)
\end{equation}

\section{Achievable Rate }

Since there is no CSIT, the total rate for each user $k$, in group
$G_{t}$, per time slot, setting $D={\displaystyle \sum_{j=1}^{K}}d_{j}^{2}$,
is given by:
\begin{equation}
R_{k}=\frac{1}{T}\log\left(1+\frac{P_{T}}{{\displaystyle \sum_{\overset{j\in G_{t}}{j<k}}}\left|\mathbf{H}_{k}\mathbf{v}_{t}\right|^{2}P_{j}+\sigma_{n}^{2}}\frac{d_{k}^{2}}{D}\left|\mathbf{v}_{t}^{T}\mathbf{H}_{k}\mathbf{v}_{t}\right|^{2}\right),
\end{equation}

where $k\in G_{t}$.

If only one user $k$ is active, with all other users shut down, the
achievable rate, per time slot, is given by:

\begin{equation}
R_{k}=\frac{1}{T}\log\left(1+\frac{P_{T}}{\sigma_{n}^{2}}\left|\mathbf{H}_{k}\mathbf{v}_{t}\right|^{2}\right)
\end{equation}

\begin{example}
For the example model, the achievable rate, setting $D={\displaystyle \sum_{j=1}^{5}}d_{j}^{2}$,
for every user is given by:
\end{example}
\begin{equation}
R_{1}=\frac{1}{2}\log\left(1+\frac{P_{T}}{\sigma_{n}^{2}}\frac{d_{1}^{2}}{D}\left|\mathbf{v}_{1}^{T}\mathbf{H}_{1}\mathbf{v}_{1}\right|^{2}\right)
\end{equation}

\begin{equation}
R_{2}=\frac{1}{2}\log\left(1+\frac{P_{T}}{\sigma_{n}^{2}}\frac{d_{2}^{2}}{D}\left|\mathbf{v}_{2}^{T}\mathbf{H}_{2}\mathbf{v}_{2}\right|^{2}\right)
\end{equation}

\begin{equation}
R_{3}=\frac{1}{2}\log\left(1+\frac{P_{T}}{\left(\left|\mathbf{H}_{3}\mathbf{v}_{1}\right|^{2}P_{1}\right)+\sigma_{n}^{2}}\frac{d_{3}^{2}}{D}\left|\mathbf{v}_{1}^{T}\mathbf{H}_{3}\mathbf{v}_{1}\right|^{2}\right)
\end{equation}

\begin{equation}
R_{4}=\frac{1}{2}\log\left(1+\frac{P_{T}}{\left(\left|\mathbf{H}_{4}\mathbf{v}_{2}\right|^{2}P_{2}\right)+\sigma_{n}^{2}}\frac{d_{4}^{2}}{D}\left|\mathbf{v}_{2}^{T}\mathbf{H}_{4}\mathbf{v}_{2}\right|^{2}\right)
\end{equation}

\begin{equation}
R_{5}=\frac{1}{2}\log\left(1+\frac{P_{T}}{\left(\left|\mathbf{H}_{5}\mathbf{v}_{1}\right|^{2}\left(P_{1}+P_{3}\right)\right)+\sigma_{n}^{2}}\frac{d_{5}^{2}}{D}\left|\mathbf{v}_{1}^{T}\mathbf{H}_{5}\mathbf{v}_{1}\right|\right)
\end{equation}

\section{Overview of results}

Our simulations were based on the example model already described.
The statistical model chosen was i.i.d. Rayleigh and our input symbols
were QPSK modulated. Maximum-Likelihood (ML) detection was performed
in the end of the decoding stage. The total transmit power was considered
as 40W (a typical value for transmit power in macrocells for 4G systems),
and therefore $a$, a constant determined by power considerations
in (5) and (6), is given by $a=\sqrt{40}$. Moreover, simulations
were performed for $100-500$ frames, with each frame consisting of
$6144$ bits.

\subsection{Degrees of Freedom (DoF)}

In {[}6{]}, with the TIM scheme, the DoF that can be achieved for
every user are 0.5 DoF, i.e. one message sent over two time slots.
In {[}8{]}, with the NOMA scheme, 1 DoF can be achieved for every
user. Introducing the hybrid scheme, results in optimal DoF for the
SISO BC channel in the cell, i.e. 
\begin{equation}
DoF_{total}=K/T
\end{equation}

\begin{figure}
\begin{centering}
\includegraphics[width=0.9\columnwidth]{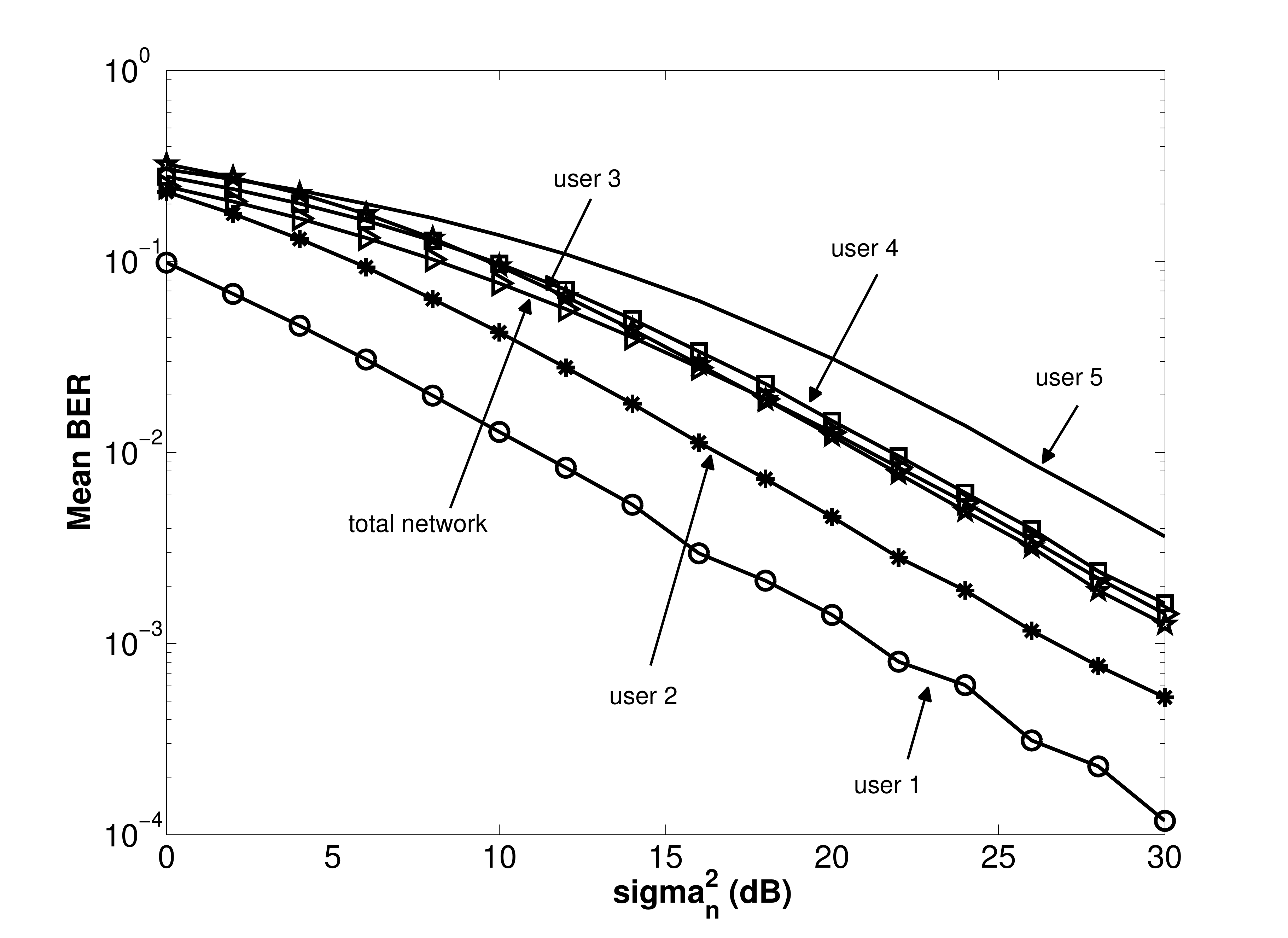}
\par\end{centering}

\caption{BER performance of the total network and every user separately. the
closer a user is to the basestation, the better its BER performance
is.}
\end{figure}

\begin{figure}
\begin{centering}
\includegraphics[width=0.9\columnwidth]{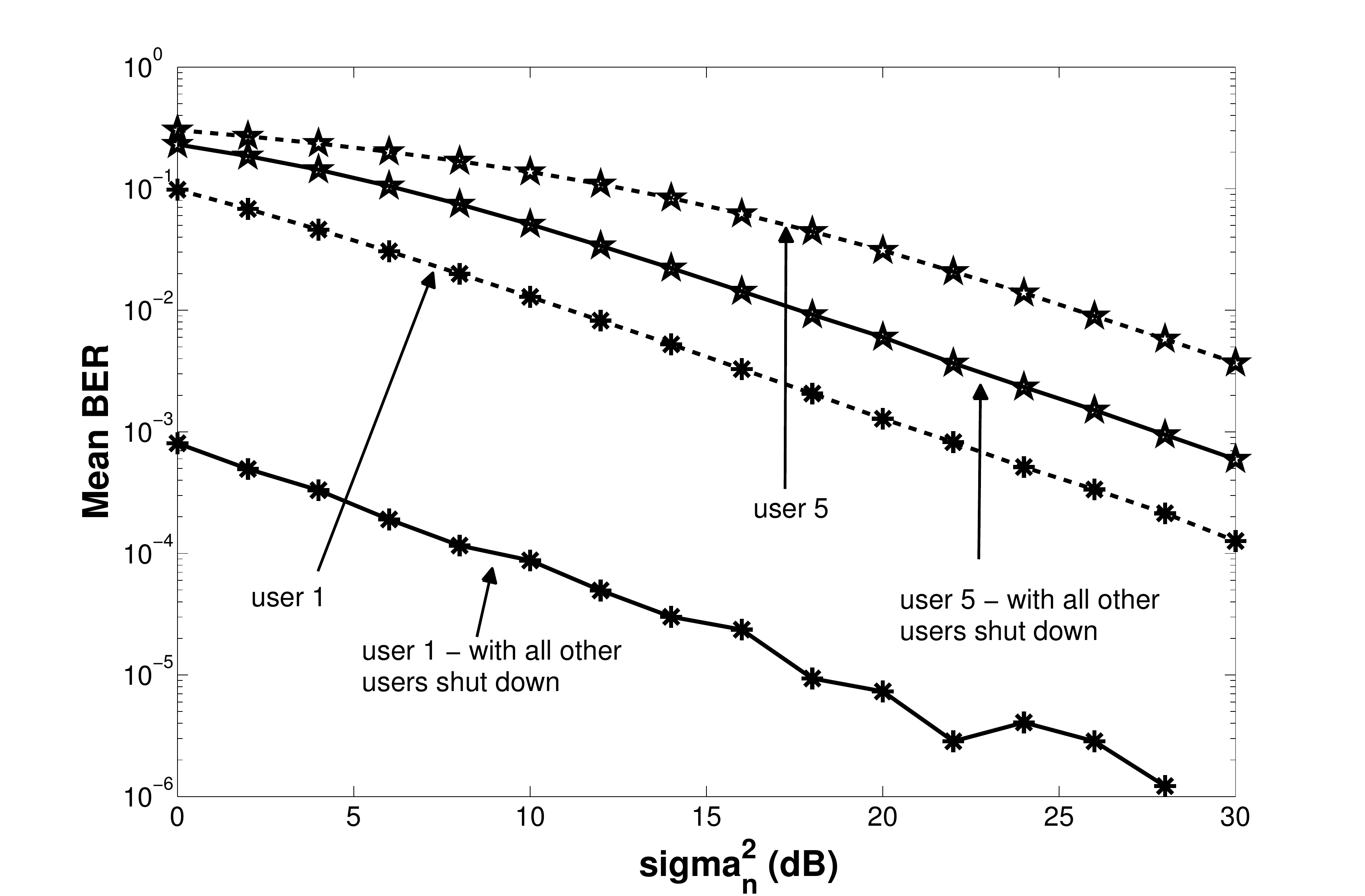}
\par\end{centering}

\caption{BER performance of every user for the hybrid scheme compared to the
one if all other users were shut down. For all users performance is
better, when only one user is active. The closer the user is to the
basestation the less the improvement is on their BER performance,
when shutting down all other users.}
\end{figure}

\subsection{Bit Error Rate (BER) Performance}

First of all, the BER performance of our example model was investigated.
Based on our findings, the distance of every user $k$ from the transmitter
is a key feature that determines the BER performance of every user.
In Figure 2, it can be observed that users who are closer to the basestation,
like users $1$ and $2$ have a better performance than users who
are far from the basestation, like users $4$ and $5$. 

Moreover, in Figure 3, a comparison between the BER performance of
users shown in Figure 2, and the performance they would achieve if
all other users were shut down, is shown. For matters of simplicity,
only users 1 and 5 are studied, as the performances of the remaining
users lie in between. As it can be observed, generally, BER performances
are better when only one user is active. Furthermore, the closer a
user is to the transmitter, the less improvement we observe in their
performance, in the case where all other users are inactive.

\begin{figure}
\begin{centering}
\includegraphics[width=0.9\columnwidth]{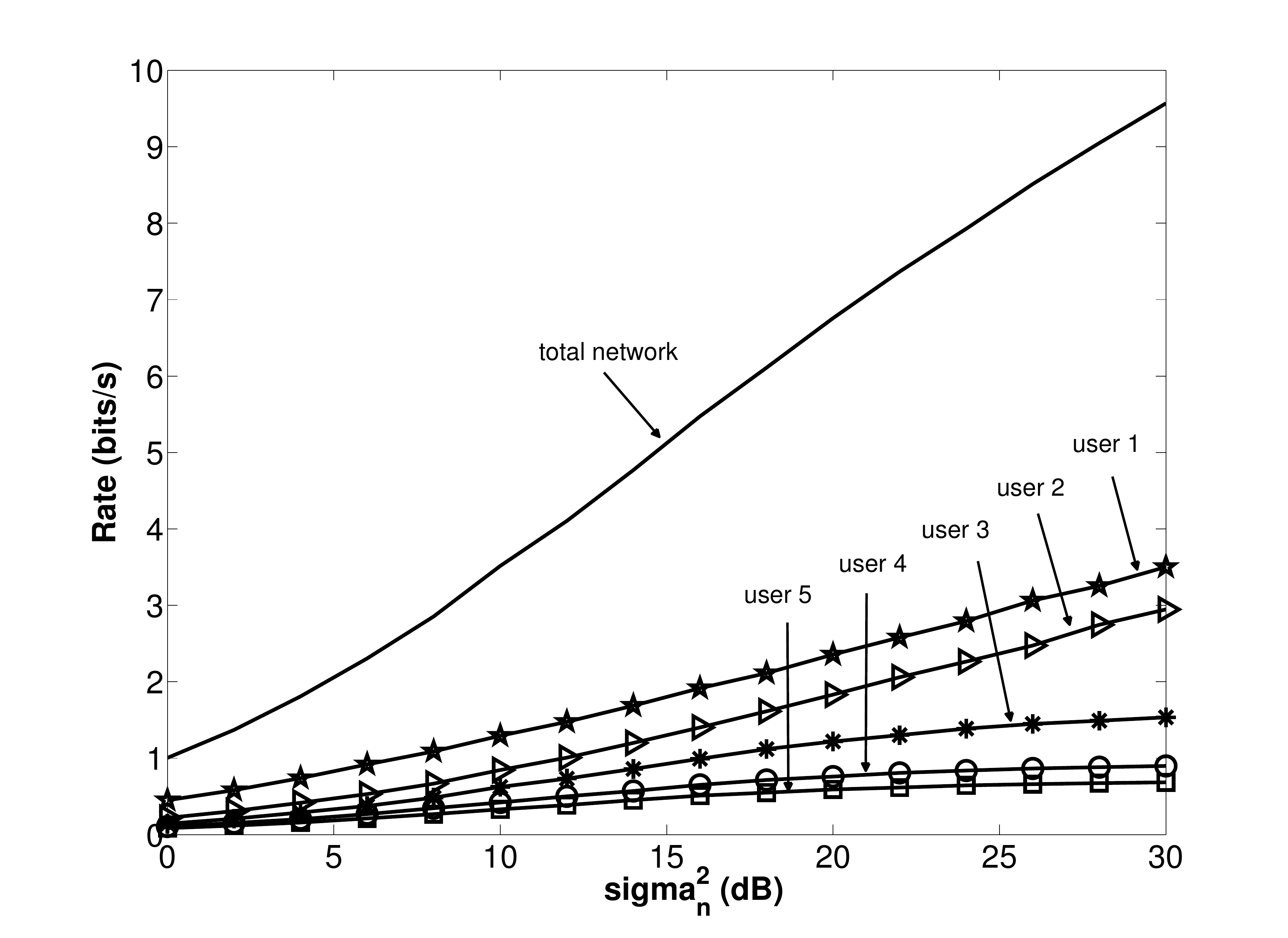}
\par\end{centering}

\caption{Rate performance of the total network and every user separately. The
closer a user is to the basestation, the better its rate performance
is.}
\end{figure}

\begin{figure}
\begin{centering}
\includegraphics[width=0.9\columnwidth]{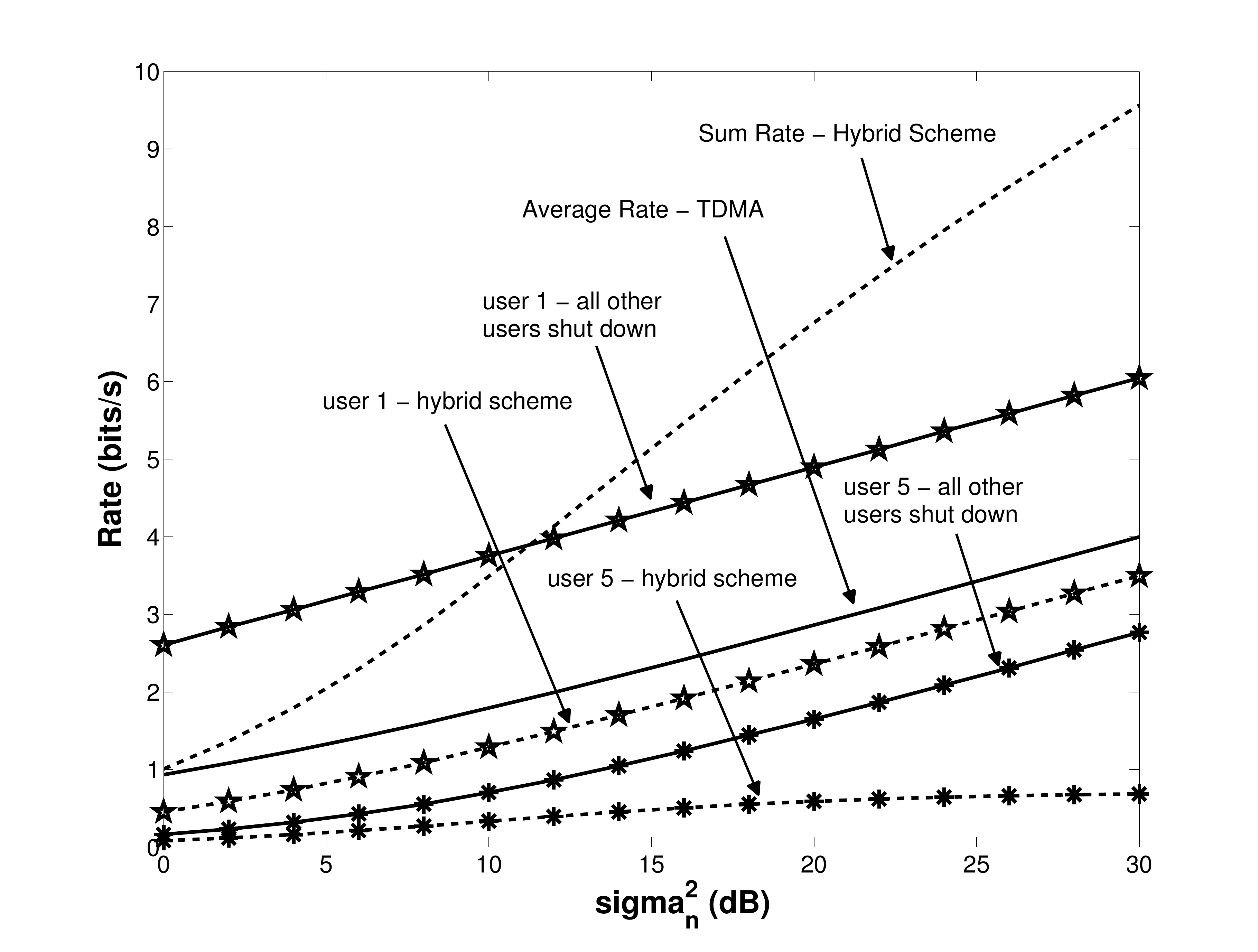}
\par\end{centering}

\caption{Rate performance of every user for the hybrid scheme compared to the
one if all other users were shut down. For all users, their rate would
be better if other users were shut down, when compared to their rate
in the hybrid scheme. However, the sum rate for the hybrid scheme
is better, for high SNRs, than the rate of user 1, when all other
users are shut down, implying the gain in terms of sum rate the hybrid
scheme provides.}
\end{figure}

\begin{figure}
\begin{centering}
\includegraphics[width=0.9\columnwidth]{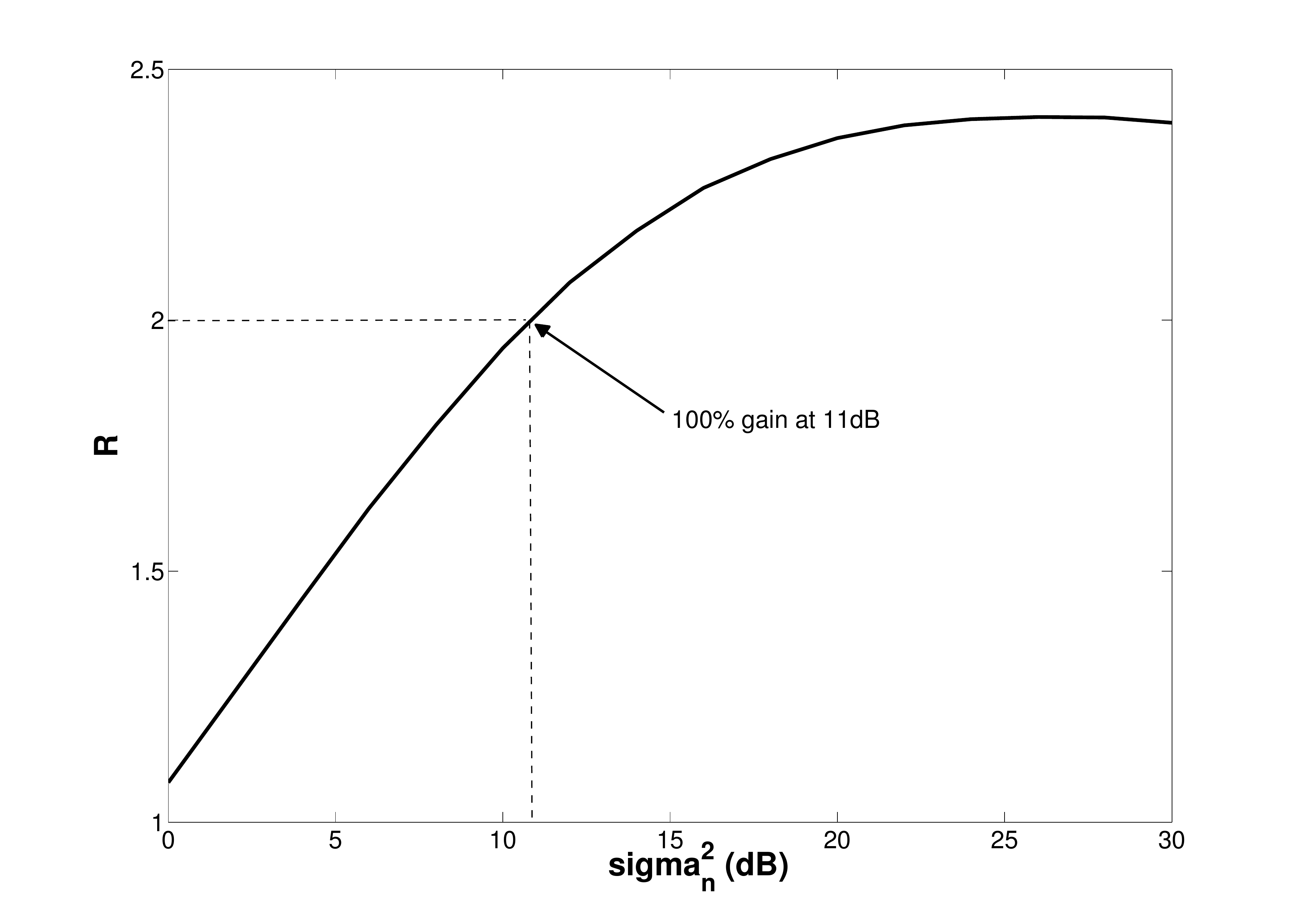}
\par\end{centering}

\caption{Ratio of sum rate of hybrid scheme over sum rate of TDMA\uline{.}
For SNR values greater than 11dB the gain, of employing the hybrid
scheme, is higher than 100\%.}
\end{figure}

\subsection{Rate Performance}

The rate of the network will be a function of the user's distance
from the basestation and the amount of interference considered as
noise, if any, as shown in (21). In Figure 4, it can be observed that
the rate decreases with the distance of the user from the transmitter
and the amount of interference considered as noise. In particular,
user $1$, who is the closest to the basestation and manages all interference
during the decoding stage, achieves the best rate performance. On
the contrary, user $5$, who is the furthest from the basestation
and considers all ``intra-group'' interference as noise, achieves
the worst performance. 

Furthermore, Figure 5 depicts a comparison between the rate for every
user shown in Figure 4, and the rate they would achieve if all other
users were shut down, as given by (22). Again, for matters of simplicity
only the cases of users 1 and 5 are shown. As it can observed, rate
performances are better when only one user is active. The most important
and interesting observation is that the sum rate for the hybrid scheme,
for high SNR values, is better than the sum rate of TDMA, proving
the gain in terms of rate that the employment of the hybrid scheme
results in. Finally, this gain is depicted clearly in Figure 6, where
the value of the ratio
\begin{equation}
\mathrm{R}=\mathrm{\frac{R_{HS}}{R_{TDMA}}},
\end{equation}

where $\mathrm{R_{HS}}$ is the sum rate of the hybrid scheme and
$\mathrm{R_{TDMA}}$ is the sum rate for TDMA, is studied for a range
of SNR values. For SNR values greater than 11 dB, the performance
of the hybrid scheme achieves at least double the rate that would
be achieved by TDMA.

\section{Summary}

Overall, this paper introduces a novel hybrid scheme that can be employed
in the SISO BC of a cell, with $K$ users divided into $T$ groups.
The hybrid scheme combines basic principles of the TIM and NOMA schemes,
by treating ``inter-group'' interference and ``intra-group'' interference
separately and by a different method. Moreover, the employment of
TIM in the cases where users' gains do not differ much, solves performance
issues that were faced by NOMA. Furthermore, the system's complexity
is reduced, providing flexibility, when compared to the NOMA scheme,
without decreasing the rate performance that the system would have
if NOMA was only applied. In general, the employment of the proposed
scheme results in high data rates, very good BER performance, and
reduced system overhead (due to the absence of CSIT requirement).
Most interestingly, for high SNRs, the total sum rate is higher than
the sum rate for TDMA, proving the gain in terms of sum rate the hybrid
scheme results in. The non-complex concept of the hybrid TIM-NOMA
scheme introduced in this paper, suggests that it could be employed
in dense networks, and potentially in heterogeneous networks once
certain adjustments in the algorithm are made.

\section{Acknowledgements}

This work was supported by NEC; the Engineering and Physical Sciences
Research Council {[}EP/I028153/1{]}; and the University of Bristol.
The authors thank Simon Fletcher \& Patricia Wells for useful discussions.


\begin{thebibliography}{10}
\bibitem{  }M. Maddah-Ali, A. Motahari, A. Khandani, ``Communications
over X channel: Signaling and performance analysis'', in \emph{Tech.
Report}, \emph{UW-ECE-2006-12}, University of Waterloo, July 2006.

\bibitem{key-1}V.R. Cadambe, S.A. Jafar, ``Interference Alignment
and Degrees of Freedom of the K-User Interference Channel'', \emph{IEEE
Trans. Inf. Theory}, vol. 54, no. 8, pp. 3425-3441, Aug. 2008.

\bibitem{google-query-cost} T. Goum, C. Wang, S.A. Jafar, ``Aiming
Perfectly in the Dark-Blind Interference Alignment Through Staggered
Antenna Switching'', \emph{IEEE Trans. Signal Process}., vol. 59,
no. 6, pp. 2734-2744, June 2011.

\bibitem{epc} S.A. Jafar, ``Blind Interference Alignment'', \emph{IEEE
J. Sel. Topics Signal Proces}., vol. 6, no. 3, pp. 216-227, June 2012.

\bibitem{key-5}S.A. Jafar, ``Elements of Cellular Blind Interference
Alignment-Aligned Frequency Reuse, Wireless Index Coding and Interference
Diversity'', arXiv:1203.2384v1, Mar. 2012.

\bibitem{key-3}S.A. Jafar, ``Topological Interference Management
through Index Coding'', \emph{IEEE Trans. Inf. Theory}, vol. 60,
no. 1, pp. 529-568, Jan. 2014.

\bibitem{key-3-1}H. Sun, S.A. Jafar, ``Topological Interference
Management with Multiple Antennas'',\emph{ IEEE ISIT}, Honolulu,
June-July 2014.

\bibitem{key-3-1-1-1}Y. Saito, Y. Kishiyama, A. Benjebbour, T. Nakamura,
A. Li, K. Higuchi, ``Non-Orthogonal Multiple Access (NOMA) for Cellular
Future Radio Access'',\emph{ IEEE} \emph{VTC Spring}, Dresden, June
2013.

\bibitem{key-3-1-1-2}A. Benjebbour, Y. Saito, Y. Kishiyama, A. Li,
A. Harada, T. Nakamura, ``Concept and Practical Considerations of
Non-Orthogonal Multiple Access (NOMA) for Future Radio Access'',\emph{
IEEE} \emph{ISPACS}, Naha, Nov. 2013.

\bibitem{key-6}Z. Ding, Z. Yang, P. Fan, H.V. Poor, ``On the Performance
of Non-Orthogonal Multiple Access in 5G Systems with randomly Deployed
Users'', arXiv:1406.1516v1, June. 2014.

\bibitem{key-7}Z. Ding, M. Peng, H.V. Poor, ``Cooperative Non-Orthogonal
Multiple Access in 5G Systems'', arXiv:1410.5846v1, Oct. 2014.

\bibitem{key-8}Z. Ding, P. Fan, H.V. Poor, ``Impact of user pairing
on 5G Non-Orthogonal Multiple Access'', arXiv:1412.2799v1, Dec. 2014.\end{thebibliography}
\end{document}